\documentclass[10pt,letterpaper]{article}

\usepackage{graphicx}
\usepackage{eepic}
\usepackage{opex3}

\begin{document}

\newcommand{\vect}[1]{{\bf #1}}

\title{Characterization of anisotropic nano-particles by using
depolarized dynamic light scattering in the near field}

\author{D. Brogioli $^1$, D. Salerno $^1$, V. Cassina $^1$, 
S. Sacanna $^2$, A. P. Philipse $^2$, F. Croccolo $^3$, and F. Mantegazza $^1$}

\address{$^1$
Dipartimento di Medicina Sperimentale, 
Universit\`a degli Studi di Milano - Bicocca,
Via Cadore 48,
Monza (MI) 20052,
Italy.
}
\address{$^2$
Van 't Hoff Laboratory for Physical and Colloid Chemistry,
Debye Research Institute,
Utrecht University,
H.R. Kruytgebouw, N-706,
Padualaan 8, 3584 CH Utrecht,
The Netherlands
}
\address{$^3$
Dipartimento di Fisica "G. Occhialini" and PLASMAPROMETEO,
Universit\`a degli Studi di Milano - Bicocca,
Piazza della Scienza 3,
Milano (MI) 20126,
Italy
}

\email{dbrogioli@gmail.com}

\begin{abstract}
Light scattering techniques are widely used in many fields of condensed and soft matter 
physics. Usually these methods are based on the study of the scattered light in the far 
field. Recently, a new family of near field detection schemes has been developed, mainly 
for the study of small angle light scattering. These techniques are based on the detection of the
light intensity near to the sample, where light scattered
at different directions overlaps but can be distinguished by Fourier transform analysis. 
Here we report for the first time data obtained 
with a dynamic near field scattering instrument, measuring both polarized
and depolarized scattered light. Advantages of this procedure over the traditional
far field detection include the immunity to stray light problems and the possibility to obtain 
a large number of statistical samples for many different wave vectors in a single instantaneous measurement. 
By using the proposed technique we have measured
the translational and rotational diffusion coefficients of rod-like colloidal particles.
The obtained data are in very good agreement with the data acquired with a
traditional light scattering apparatus.
\end{abstract}


\section{Introduction}

Light scattering has been extensively used for many years to measure statistical properties of
samples coming from many different sources, ranging from soft matter (e.g. polymers, colloids,
gels, emulsions) to biology (e.g. blood cells, vesicles); 
see for instance the recent review \cite{scheffold2007} and references therein. 
Several different experimental techniques have been developed and among 
them we can mention small angle light scattering \cite{ferri1997, cipelletti1999},
two-colour dynamic light scattering \cite{schatzel1991, pusey1999}, fiber optical
quasi elastic light scattering \cite{thomas1989},
diffusing wave spectroscopy \cite{pine1988}, and particle optical tracking \cite{grier2007}.

Traditional techniques for the detection of the scattered light 
operate in the so called far field regime
(scattering in far field, SIFF). In this case, the intensity of the light scattered at 
different directions is measured by placing a sensor far from the sample, 
where beams with different 
directions are also spatially separated. Usually, in the small angle scattering range, the far
field condition is obtained by locating the detecting sensor (a CCD or other pixilated sensor)
in the focal plane of a lens positioned in front of the sample \cite{cipelletti1999}.
However, many techniques have recently been developed to measure the scattered light 
in the near field (scattering in near field, SINF)
i.e. with the detector placed very close to the sample. In this case the image is obtained as a
superposition of beams pointing in several different directions. 
The SINF technique family includes near field scattering
\cite{brogioli2002, brogioli_phd}, based on the ``near field speckles'' concept 
\cite{giglio2000, giglio2001, brogioli_phd},
shadowgraph \cite{wu1995,trainoff2002} and schlieren \cite{brogioli2003}.

In this way it is possible to obtain not only static measurements of the scattered intensity but also dynamic data 
related to the diffusion coefficients of the scattering objects. 
Both static and dynamic data are simultaneously 
measured at several different angles.
Dynamic scattering information has been extracted via several slightly
different SINF techniques; among them, the exposure time dependent spectrum (ETDS)
technique \cite{oh2004, brogioli2008}, closely related to ``speckle visibility spectroscopy'' \cite{durian2005},
and the differential algorithm \cite{croccolo2006bis, croccolo2007, magatti2008, cerbino2008}.

In the present paper, we address the problem of measuring the depolarized light
scattering; i.e. scattering with polarization different from the impinging beam, 
usually due to the presence of anisotropic scatterers within the sample. 
The interest in this problem is widespread because dynamics of the
depolarized scattering from anisotropic colloids is connected to their rotational
diffusion, which in turn is hard to measure and object of several studies
\cite{philipse2003}.
We perform the analysis of the depolarized scattering light by simply applying a
polarizer/analyzer scheme to a traditional SINF apparatus: 
this allows the transmitted beam
and the perpendicularly polarized scattered beam to interfere.
In order to test our approach, the resulting rotational diffusion coefficients 
of rod-like colloidal particles have been successfully compared with the data 
taken with a traditional SIFF apparatus. 

\section{Experimental set-up}

\begin{figure}
\begin{center}
\begin{picture}(330,180)(-100,0)
%
%
%
%
\Thicklines
\path(-100,0)(-20,0)(-20,30)(-100,30)(-100,0)
\put(-80,10){Laser}
\path(150,5)(170,25)
\put(162,10){Mirror}
\path(130,80)(190,80)(190,90)(130,90)(130,80)
\put(140,82){Sample}
\path(130,93)(157,93)
\path(163,93)(190,93)
\put(180,95){Iris}
\path(140,175)(180,175)(180,180)(140,180)(140,175)
\put(110,170){CCD}
%
%
\thicklines
\path(-30,4)(-25,4)(-25,26)(-30,26)(-30,4)
\put(-55,32){Polarizer}
\path(60,4)(65,4)(65,26)(60,26)(60,4)
\put(40,28){Neutral filter}
\path(145,33.980762114)(145,46.019237886)
\path(175,33.980762114)(175,46.019237886)
\put(160,72){\arc{60}{1.0471973335}{2.0943946664}}
\put(160,8){\arc{60}{-2.0943946664}{-1.0471973335}}
\put(180,35){f=-5cm}
\path(145,108)(145,112)
\path(175,108)(175,112)
\put(160,133.98076211){\arc{60}{1.0471973335}{2.0943946664}}
\put(160,86.019237886){\arc{60}{-2.0943946664}{-1.0471973335}}
\put(95,115){Microscope}
\put(95,105){objective}
\path(149,140)(171,140)(171,145)(149,145)(149,140)
\put(100,140){Analyzer}
%
%
\thinlines
\path(-40,17)(162,17)(162,40)(180,80)
\path(-40,13)(158,13)(158,40)(140,80)
\path(160,93)(170,110)(160,175)(150,110)(160,93)
\end{picture}
\end{center}
\caption{Scheme of the optical set-up. The He-Ne laser, containing the polarizer,
generates a collimated laser beam, which is
attenuated by a neutral filter and bent upwards by a mirror. The beam is
expanded by means of a negative focal length lens, making it diverge slightly
before going through the sample cell. Scattered light is acquired in the near field
together with transmitted light, through a 40x microscope objective, which 
conjugates a plane close to the sample onto the CCD sensor. 
Between the microscope objective and the CCD sensor, 
a rotating polarizer, which can be oriented at any angle, is inserted as analyzer.
}
\label{fig_setup}
\end{figure}
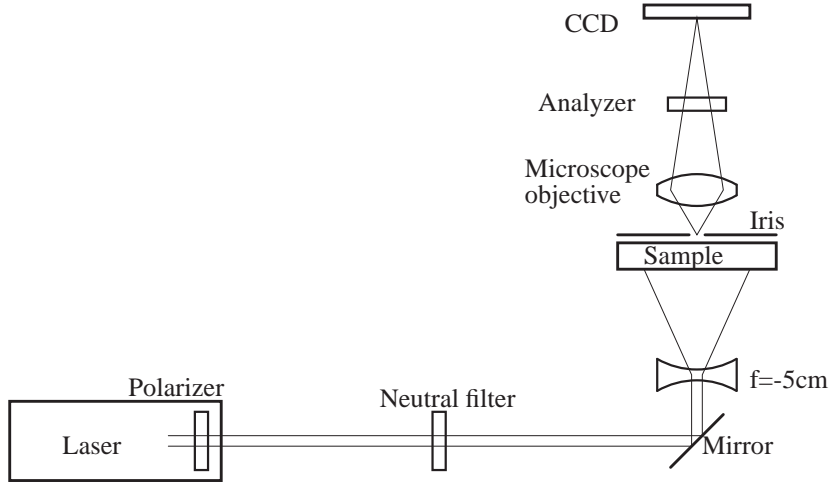
The experimental set-up shown in Fig. \ref{fig_setup}, is similar to the one
described in \cite{brogioli2008}.
and basically consists of a standard microscope
equipped with a low-speed camera. The illuminating
lamp has been replaced by a 10mW He-Ne laser (Nec), enlarged
to a diameter of $4\mathrm{mm}$ at $1/e$ by a negative lens. The laser intensity
is attenuated by a neutral filter, selected by a filter wheel.

The sample is placed in a glass cell with optical path 1 mm,
made of two microscope cover slips, spaced by small glass strips
cut from a microscope slide, and glued with silicone rubber.
In order to reduce unwanted reflections, an iris with a 1 mm-diameter
aperture was placed over the top cover slip. The optical system
consists of a plan-achromatic 40x objective (Optika Microscopes,
FLUOR), with 0.65 numerical aperture, 160mm focal distance
and 0.17mm working distance; its focus lies on the iris plane,
about 0.2mm outside the cell. Images are acquired with
a CCD camera (Andor Luca), whose sensor is 658 x 496 pixels.
The camera maximum frame rate is 30 frames per second, with
a minimum exposure time of 0.5ms. The sensor is placed at 160
mm from the microscope objective, so that it collects images
directly with a magnification of 40x. The CCD sensor images
an area of $200 x 150\mathrm{\mu m}^2$. The diameter of the 
illuminated area of the sample is about 4mm; the imaged area
sees a region of illuminated sample for every direction inside the
numerical aperture of the objective; this ensures that
near field detection of the light can be performed over the whole
accepted scattering angles \cite{brogioli2003, magatti2008}.

The polarizer/analyzer scheme is simply constituted by an analyzer 
(Techspec Linear Polarizer, Edmund Optics)
placed between the objective and the camera, which
can be rotated at any angle $\vartheta$ with respect to the
transmitted beam. The polarizer is already included in the laser cavity,
resulting in a very good linear polarization of the out-coming light.

Each experiment consists in collecting the images, taken by the CCD, at
different exposure times 
and at different polarizing angles $\vartheta$ between the polarizer and the analyzer.
In order to change the exposure time, we first select a neutral filter, and if needed
adjust the exposure time so that the acquired images have an average intensity
corresponding to half the camera dynamic range, thus ensuring an optimal digitization.
One hundred raw images are acquired, at a frame rate of 1 Hz, slowly enough to
ensure that all the images are uncorrelated. The average over the 
raw images is then subtracted from each image, in order to get rid of stray light
and beam inhomogeneities. 
Data are collected at two different polarizing angles,
namely $\vartheta=0$ (polarized VV scattering) and
$\vartheta \approx 90^{\circ}$ (polarized VV + depolarized VH scattering, see below).

\section{Materials and sample characterization}

We used two rod-like samples: a polytetrafluoroethylene colloid (PTFE) and
akaganeite (iron oxide) rods (DR1).

The PTFE colloidal dispersion is a diluted aqueous suspension of submicronic 
fluorinated particles made of crystalline polytetrafluoroethylene (PTFE) which have
been extensively studied in the past \cite{bellini1995, degiorgio1994, piazza1989}.
The particles were kindly donated by Solvay-Solexis, Bollate (Milano), Italy. 
The particles have been produced by polymerizing fluorinated monomers at high pressure
in the presence of a fluorinated surfactant, using a free-radical initiator
\cite{bellini1995, degiorgio1994}.
The particles have a rodlike shape with a polydispersity in
the linear dimension of about 15\%. 
The particles have an
internal crystalline structure that makes them intrinsically
birefringent \cite{bellini1988}, with the fast axis
in the direction 
of the symmetry axis of the particle; the corresponding
internal birefringence is
about $\Delta n = 0.04$, and their average refractive index is $n=1.37$.

In order to characterize the PTFE sample and in particular its size and shape, 
we have studied both the polarized
(VV) and depolarized (VH) light scattering components with
a traditional dynamic SIFF apparatus (or dynamic light scattering DLS).
SIFF measurements are aimed at measuring the intensity of the beams scattered 
with transferred wave vector $\vect{Q} = \vect{K}_S - \vect{K}_I$, 
$\vect{K}_S$ and $\vect{K}_I$ being
the wave vectors of the scattered and incident beams, respectively. 
Characterization measurements have been
performed at scattering angles of $15^{\circ}$ and $90^{\circ}$ 
using a $659\mathrm{nm}$ diode laser. 
The measured intensity autocorrelation functions for both the VV 
and VH components are with a good approximation monoexponential. 
Indeed, the field autocorrelation functions $C_E^{VV}$ and $C_E^{VH}$ expected for scattering
of a monodisperse, anisotropic colloidal sample are monoexponential \cite{berne2000}:
\begin{eqnarray}
C_E^{VV}\left(Q,t\right) \propto e^{-t/\tau_{VV}}\\
C_E^{VH}\left(Q,t\right) \propto e^{-t/\tau_{VH}} ,
\end{eqnarray}
where
\begin{eqnarray}
\tau_{VV} = \frac{1}{D_T Q^2}
\label{eq_tau_VV}
\\
\tau_{VH} = \frac{1}{D_T Q^2 + 6\Theta}
\label{eq_tau_VH}
\end{eqnarray}
are the heterodyne decay times, 
$D_T$ is the translational diffusion coefficient, and $\Theta$ is the rotational
diffusion coefficient of the scattering objects. 
It is worth pointing out that while the translational time constant changes with the wave vector, 
the rotational one is independent of it. The predicted
intensity autocorrelation functions can be simply obtained by the Siegert
relation \cite{goodman2005} as follows:
\begin{eqnarray}
C_I^{VV}\left(Q,t\right) \propto \beta e^{-2 t/\tau_{VV}}+1\\
C_I^{VH}\left(Q,t\right) \propto \beta e^{-2 t/\tau_{VH}}+1,
\end{eqnarray}
where $\beta$ is a constant depending on the detector geometry. Fitting the above formulae to our
experimentally measured autocorrelation functions allows the evaluation of the 
translational and rotational diffusion coefficients of PTFE:
$D_T=1.5\cdot 10^{-12}\mathrm{m}^2/\mathrm{s}$; $\Theta=24.5\mathrm{s}^{-1}$.
In turn, from these values, under the hypothesis
that the particles are prolate ellipsoids, the particle size can be estimated. 
Accordingly the obtained major axis is
$640\mathrm{nm}$; while the minor one is $150\mathrm{nm}$, thus providing a form factor of about 4,3:1.
The PTFE sample was diluted
to a 0.25\% volume fraction, but in order to exclude artifacts due to multiple
scattering, we tested also the
0.1\% concentration, obtaining similar results.

The DR1 sample is a suspension of iron oxide akaganeite needles.
Akaganeite ($\beta-\mathrm{FeOOH}$) rods were prepared following a modified synthesis originally 
developed by Sugimoro et al. \cite{sugimoto1992}.
Briefly, a highly condensed ferric hydroxide gel ($\mathrm{Fe\left(OH\right)_3}$) 
was aged at $100^{\circ}\mathrm{C}$ for 48 hours inside a sealed Pyrex bottle. The resulting precipitate was then 
quenched at room temperature, washed and resuspended in deionized water. This yielded a 
stable suspension of akaganeite needle-like particles with a polydispersity, determined by 
TEM analysis, of about 30\% (more details can be found in \cite{sacanna2007}).
The VV and VH intensity autocorrelation functions
obtained with traditional SIFF apparatus on the DR1 sample are 
monoexponential for the VV component, but stretched exponential 
for the VH component:
\begin{equation}
C_I^{VH}\left(Q,t\right) \propto \beta e^{-2 \left(t/\tau_{VH}\right)^{\alpha}}+1,
\label{eq_stretched_exponential}
\end{equation}
The SIFF fitting parameters are: $D_T = 2.1\cdot 10^{-12}\mathrm{m}^2/\mathrm{s}$;
$\Theta = 27.8\mathrm{s}^{-1}$; $\alpha = 0.523$.
Such a value of the stretching exponent $\alpha$ confirms that the sample
has a large degree of polydispersity as observed in \cite{sacanna2007}.
Given the non exponential decay of the correlation functions, 
it is difficult to have a precise evaluation of the size of the DR1 rods 
which have to be compared with average values obtained by TEM:
major axis $150\mathrm{nm}$, minor axis $10\mathrm{nm}$.

\section{Methods}

In this section we briefly outline the principles of the SINF technique
\cite{brogioli2002} and its extension to depolarized measurements.

Our set-up is based on an optical scheme belonging to the SINF family,
which allows the detection of the scattering light by imaging a sample area in the near field.
In particular, the present apparatus is an heterodyne SINF setup \cite{brogioli2002},
where both the transmitted and scattered beams are collected.
The imaged area is illuminated by the transmitted light, which is collected by the detector together
with the much weaker beams scattered by the colloidal particles. 
The collecting wave vector range is determined by the objective's numerical aperture. 
The use of an objective instead of a lens allows a larger range of available $Q$. 
The obtained image is recorded with a pixelated sensor, digitalized and then Fourier transformed by 
standard software packages. The power spectrum $S\left(Q\right)$ is then obtained as
the mean square modulus of the Fourier transform of the image.
The fundamental idea underlying SINF techniques is that each scattered beam, 
with transferred wave vector $\vect{Q}$, 
generates exactly one Fourier mode of the image,
with two-dimensional wave vector $\vect{q}$. Thus a Fourier transform of the image
readily gives the scattered fields, and the power spectrum gives the scattered intensities.
The quantitative relation between the magnitudes of the scattering wave vector 
$\vect{Q}$ and the image wave vector $\vect{q}$ is:
\begin{equation}
Q\left(q\right)=\sqrt{2}K\sqrt{1-\sqrt{1-\left(\frac{q}{K}\right)^2}}
\end{equation}
where $K=K_I=K_S$ is the light wave vector in the medium. The 
approximation $Q \approx q$ holds in our case, since scattering angles and therefore image wave vectors $q$
are small. The relation between image power
spectrum $S\left(q\right)$ and scattered intensity $I\left(Q\right)$ is:
\begin{equation}
S\left(Q\right) = T\left(Q\right) I\left(Q\right)+ B\left(Q\right)
\end{equation}
where $T\left(Q\right)$ is a transfer function heavily depending on the instrumental
setup (see for example \cite{trainoff2002} and \cite{brogioli2002}) and 
$B\left(Q\right)$ is the instrumental noise mainly due to the electronics.
We abuse the notation by using  
$S\left(Q\right)$ for $S\left[q\left(Q\right)\right]$,
that is, we express the image power spectra $S\left(q\right)$ as
a function of the 
corresponding transferred wave vector $Q$, which holds only if $q<<k$.

The VV SINF signal is obtained by placing the analyzer along the direction determined by the transmitted beam
polarization, that is with $\vartheta=0$. 
Conversely, in order to observe the interference between the transmitted 
beam and the horizontally polarized scattered
beams (VH), we rotate the analyzer at an angle $\vartheta$ close
to $90^{\circ}$. The angle cannot be exactly $90^{\circ}$ (crossed polarizers), nor too 
close to $90^{\circ}$, because in order to maintain the heterodyne detection scheme it is
necessary that the transmitted beam, attenuated by the
nearly crossed polarizer, be much stronger than the scattered beams \footnote{The scattered beams generate a 
speckle field. The intensity distribution of an homodyne image has an exponential
decay; if the speckle field interfere with a much stronger transmitted beam, the image
is heterodyne, and the intensity approaches a gaussian distribution, centered at the intensity
of the transmitted beam. We use the intensity distribution of the images to check if
we are working in heterodyne regime.}. Beyond the analyzer, the attenuated transmitted beam 
has the same polarization as the (less attenuated) VH scattered 
beam, and this generates the VH SINF signal. Several angles close to $90^{\circ}$
have been tested obtaining basically similar results. Here we present data obtained at $\vartheta = 82.0^{\circ}$
being the best compromise between a good signal-to-noise ratio and the heterodyne condition.

Dynamic analysis is then performed by the so called
exposure time dependent spectra (ETDS) technique, which we recently introduced
\cite{brogioli2008}, that is, by measuring power spectra of images taken with 
different exposure times.

\section{PTFE experimental results}

\begin{figure}
\begin{center}
\begin{tabular}{p{4.0cm}p{4cm}}
\multicolumn{1}{c}{a}&\multicolumn{1}{c}{b}\\
\includegraphics[scale=0.5]{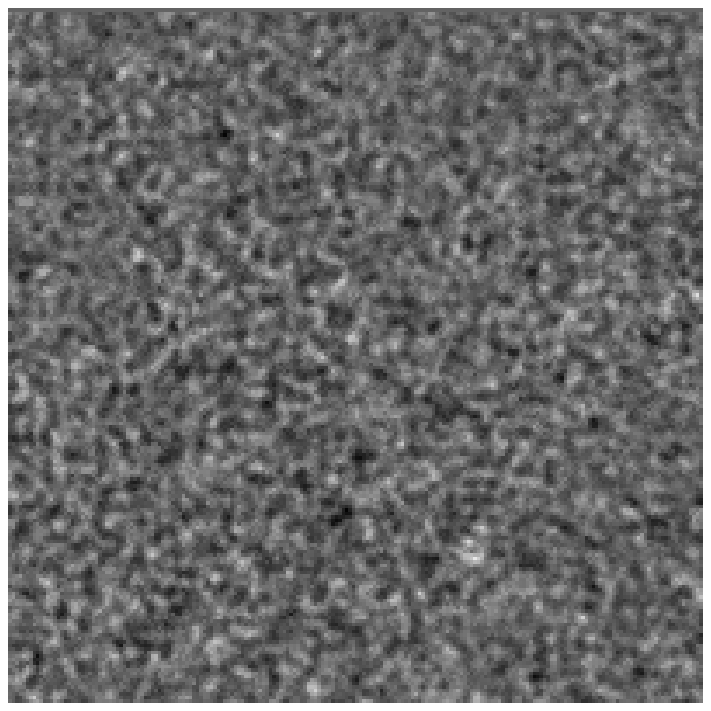}
&
\includegraphics{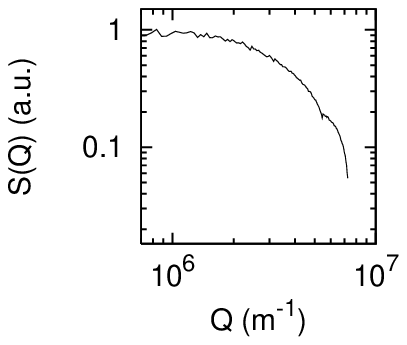}
\\
\multicolumn{1}{c}{c}&\multicolumn{1}{c}{d}\\
\includegraphics[scale=0.5]{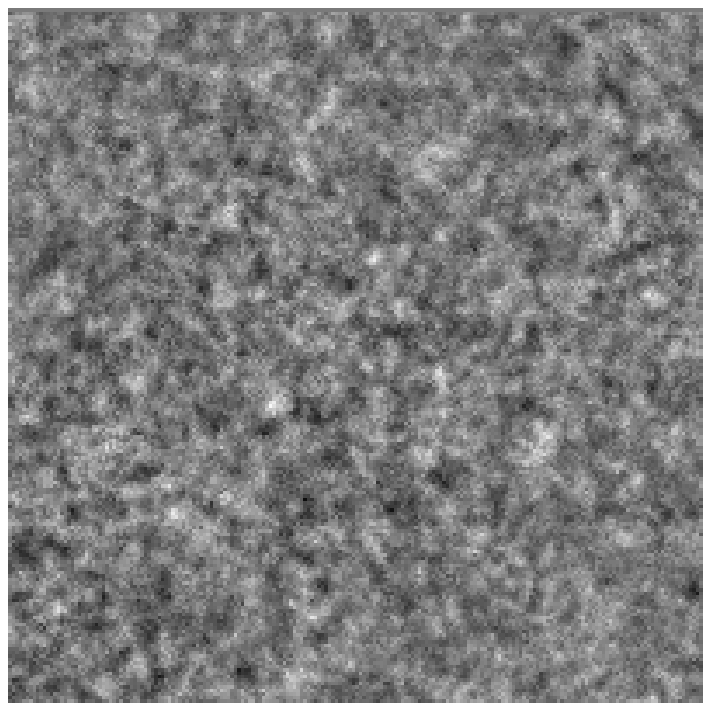}
&
\includegraphics{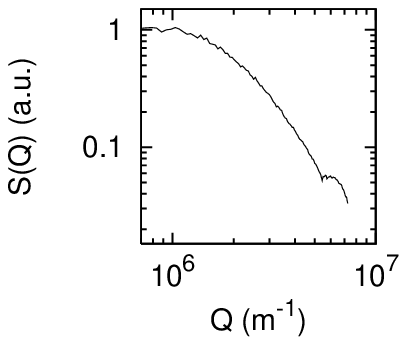}
\end{tabular}
\end{center}
\caption{Images collected by the CCD using the SINF technique (panels a and c) and their 
power spectra $S\left(Q\right)$ (panels b and d). Data taken at $\vartheta=0$ for two different
exposure times: $\Delta t=48\mathrm{ms}$ for panels a and b;
$\Delta t=497\mathrm{ms}$ for panels c and d. Data for PTFE dispersion, 0.25\% volume fraction in water. }
\label{fig_immagini_pol}
\end{figure}

Figure \ref{fig_immagini_pol} shows two typical images taken with our system, for a PTFE
sample at 0.25\% volume fraction, at two different 
exposure times. In these pictures the analyzer is at $\vartheta=0^{\circ}$, that is, aligned
with the transmitted beam.
Actually the speckle pattern is not static in time
but seems to ``boil'', as consequence of the motion of the scattering particles, which
generates changes in the beam phase.
The image taken at an exposure time $\Delta t=48\mathrm{ms}$ is nearly freezed; on the contrary,
the image at $\Delta t=497\mathrm{ms}$ is blurred by the averaging of the speckle motion. Sets
of images taken at different exposure times $\Delta t$ contain information on the dynamics
of the scattering objects.

The characteristic power spectra $S\left(Q\right)$ of the speckle images are also shown in Fig.
\ref{fig_immagini_pol}, where the graphs represent the angular average of the spectra.
The image blurring, due to large exposure time, appears as a depression, which is more evident at
large wave vectors. Actually, all the information on the dynamics of the image Fourier modes, and 
hence of the scattered fields, is
contained in the ETDS $S\left(Q,\Delta t\right)$, that 
is, in the image power spectra taken at several different exposure times \cite{brogioli2008}.

\begin{figure}
\includegraphics{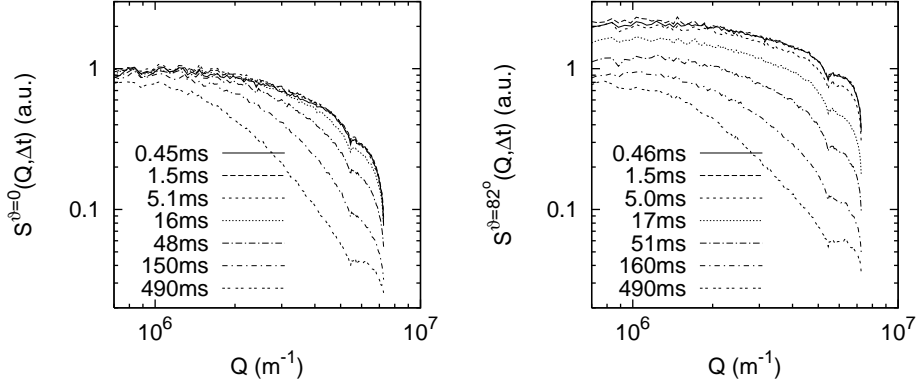}
\caption{Exposure time dependent spectra $S^{\vartheta}\left(Q,\Delta t\right)$, measured as functions 
of the scattering vector $Q$, at two different angles $\vartheta$ between polarizer and analyzer.
Left panel: data taken at different exposure time $\Delta t$ with the analyzer at $\vartheta=0$
(VV, polarized component).
Right panel: same as left panel, but with analyzer at $\vartheta = 82.0^{\circ}$ (VV + VH, sum of polarized
and depolarized components).
The keys provide the different values of the exposure time $\Delta t$.
Data taken for PTFE as in Fig. \ref{fig_immagini_pol}.
}
\label{fig_PTFE_spettro_pol_depol}
\end{figure}

Indeed, a set of ETDS, taken at different exposure times, at
$\vartheta=0$, is shown in Fig. \ref{fig_PTFE_spettro_pol_depol}.
The graph on the left refers to the VV component, with $\vartheta=0$.
It can be noticed that, as $\Delta t$ increases, the power spectra 
$S^{\vartheta=0}\left(Q,\Delta t\right)$ taken at $\vartheta=0$ decrease.
Moreover, the decrease is stronger at large wave vectors.
This can be qualitatively interpreted as a consequence of the translational diffusion of
the colloidal particles: light scattered at the transferred wave vector $Q$
has a characteristic time $\tau_{VV}$, decreasing as $Q$ increases, see Eq. (\ref{eq_tau_VV}).
For large $Q$ values, the corresponding Fourier modes oscillate
faster, and hence the averaging cancels them at shorter exposure times.
The graph on the right in Fig. \ref{fig_PTFE_spettro_pol_depol} refers to 
data measured at nearly crossed
polarizers, namely $\vartheta=82.0^{\circ}$. As apparent from the graph,
as the exposure time $\Delta t$ increases, the ETDS decreases approximately at the
same rate for all wave vector. This fact is explained if we recall that the power spectra
$S^{\vartheta=82.0^{\circ}}\left(Q,\Delta t\right)$ mainly reflects the VH scattering
dynamics. For small wave vectors $Q$, $\tau_{VH}$ scales with the rotational
diffusion of the anisotropic colloidal particles, thus being basically constant, see Eq. (\ref{eq_tau_VH}).

\begin{figure}
\begin{center}
\includegraphics{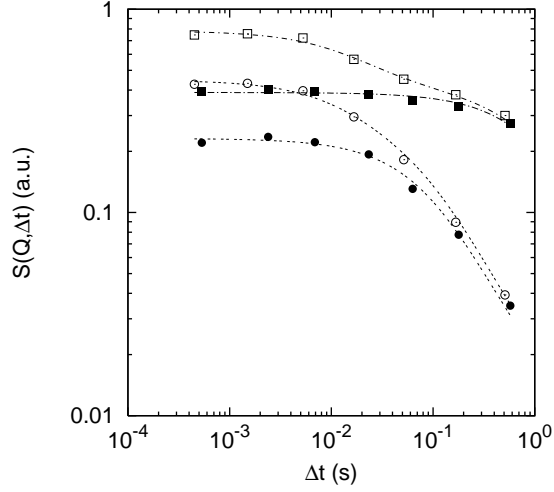}
\end{center}
\caption{
Exposure-time dependent spectra $S\left(Q,\Delta t\right)$, measured as functions
of the exposure time $\Delta t$, at two different angles $\vartheta$ between polarizer and analyzer.
Squares: $Q=1.23\cdot 10^6 \mathrm{m}^{-1}$; dots: $Q=4.42\cdot 10^6 \mathrm{m}^{-1}$.
Filled symbols: $\vartheta=0$; open symbols: $\vartheta=82.0^{\circ}$.
The lines represent the fitting curves obtained according to 
Eq. (\ref{eq_ETDS_esponenziale}), see text for details. Fitting parameters
at $Q=1.23\cdot 10^6 \mathrm{m}^{-1}$,          $\tau_{VV}=0.5\mathrm{s}$ and $\tau_{VH}=6.2\mathrm{ms}$;
at $Q=4.42\cdot 10^6 \mathrm{m}^{-1}$,          $\tau_{VV}=36\mathrm{ms}$ and $\tau_{VH}=5.5\mathrm{ms}$.
Data taken for PTFE as in Fig. \ref{fig_immagini_pol}.
}
\label{fig_PTFE_sezioni}
\end{figure}

Figure \ref{fig_PTFE_sezioni} shows graphs of the ETDS as a function of
the exposure time $\Delta t$, at two fixed $Q$ vectors: 
$Q=1.23\cdot 10^6\mathrm{m}^{-1}$ and $4.42\cdot 10^6\mathrm{m}^{-1}$,
for two different angles between the polarizer and the analyzer
$\vartheta = 0$ and $\vartheta=82.0^{\circ}$.
The data in Fig. \ref{fig_PTFE_sezioni} are obtained from the data in Fig. \ref{fig_PTFE_spettro_pol_depol},
and reflect 
the different blurring of the images at two values of $\vartheta$ and two values of $Q$.
It is apparent that the four curves show four different decay
behaviors. At small $Q$ the $S\left(Q,\Delta t\right)$ is larger than at large $Q$. Moreover,
the scattering at nearly crossed polarizers ($\vartheta=82.0^{\circ}$) approximates
the $\vartheta=0$ scattering at large exposure times. 
This is due to the fact that the rotational time constant is always smaller than the translational one.

\section{PTFE discussion}

In this section we give a quantitative assessment of the effect of 
the analyzer on the power spectra $S\left(Q\right)$ of PTFE colloidal 
particles. We consider 
the image field as the superposition of three fields: $E_0$,
the transmitted beam, and the scattered fields $E_{VV}$ and $E_{VH}$,
parallel and perpendicular to $E_0$:
\begin{equation}
\vect{E} = \vect{E}_0 + \vect{E}_{VV} + \vect{E}_{VH}
\end{equation}
Beyond the analyzer, the field is:
\begin{equation}
E =  E_0 \cos \vartheta + E_{VV} \cos \vartheta + E_{VH} \sin \vartheta
\end{equation}
The intensity is thus proportional to:
\begin{equation}
I = \left|E\right|^2 \propto
\left|E_0\right|^2 + 
\Re\left[E_{VV} E_0^*\right] +
\Re\left[E_{VH} E_0^*\right] \tan \vartheta 
\end{equation}
where $\Re$ is the real part and $^*$ is the complex conjugate and the terms $E_{VV}^2$ and $E_{VH}^2$ 
can be neglected because 
we consider the transmitted beam much stronger than scattered ones (heterodyne condition).
All the data processing is performed after normalization 
of the images' intensities at a given average intensity. Changing $\vartheta$
modifies only the relative amplitude of the VH contribution to the image, while
the VV term is left constant. For our 
samples, which scatter almost isotropically,
the cross correlations between $E_{VV}$ and $E_{VH}$ are negligible. Hence the
power spectra is the sum of the VV and VH power spectra:
\begin{equation}
S^{\vartheta}\left(Q,\Delta t\right) = 
S^{VV}\left(Q,\Delta t\right) + S^{VH}\left(Q,\Delta t\right) \tan^2 \vartheta 
\label{eq_S_VV_VH}
\end{equation}
At $\vartheta=0$, we measure the VV signal only:
\begin{equation}
S^{VV}\left(Q,\Delta t\right) =
S^{\vartheta=0}\left(Q,\Delta t\right) 
\label{eq_S_VV_esplicita}
\end{equation}
Increasing $\vartheta$, 
we get a mixture of VV and VH components.
It can be noticed that the relative intensity of the VV component with respect to 
the transmitted beam is left unchanged by modifying the analyzer position,
and it always contributes to the SINF signal by the same amount.
As $\vartheta$ approaches $90^{\circ}$, the relative amplitude of the VH component of the EDTS increases.
However, arbitrarily large values cannot be reached, because of the existence of a limit angle, 
above which the heterodyne condition is violated.

As shown in Fig. \ref{fig_PTFE_sezioni}, rotating the analyzer
towards $90^{\circ}$ has the effect of increasing the measured ETDS values.
The increment, that is the difference
between the ETDS at $\vartheta = 82.0^{\circ}$ and $\vartheta = 0$, is
proportional to the depolarized (VH) ETDS:
\begin{equation}
S^{VH}\left(Q,\Delta t\right) \propto
S^{\vartheta=82}\left(Q,\Delta t\right) - S^{\vartheta=0}\left(Q,\Delta t\right)
\label{eq_S_VH_esplicita}
\end{equation}

It has been shown that the relation between $C_E\left(Q,t\right)$ and the ETDS
$S(Q,\Delta t)$ is \cite{brogioli2008}:
\begin{equation}
S\left(Q,\Delta t\right) \propto \frac{2}{\Delta t^2} \int_0^{\Delta t}\mathrm{d}s
\left(\Delta t -s\right)C_E\left(Q,s\right)
\label{eq_ETDS}
\end{equation}
This relation allows us to interpret the ETDS data and to compare them with the 
results of the dynamic SIFF measurements.
For the case of an exponential decay $C_E\left(q,t\right) \propto exp\left[-t/\tau\right]$,
we have \cite{oh2004, durian2005, brogioli2008}:
\begin{equation}
S\left(Q,\Delta t\right) \propto f\left(\frac{\Delta t}{\tau}\right)
\label{eq_ETDS_esponenziale}
\end{equation}
where:
\begin{equation}
f\left(x\right) = \frac{2}{x^2} \left(e^{-x}-1+x\right)
\label{eq_ETDS_forma}
\end{equation}
Indeed, the ETDS for $\vartheta=0$ in Fig. \ref{fig_PTFE_sezioni} can be nicely fitted
with Eq. (\ref{eq_ETDS_esponenziale}) with $\tau_{VV}$ as fitting parameter (see filled
symbols in Fig. \ref{fig_PTFE_sezioni} and the fitting curves).

The VH ETDS has been obtained by Eq. (\ref{eq_S_VH_esplicita}) and has been 
fitted with Eq. (\ref{eq_ETDS_esponenziale}), with $\tau_{VH}$ as fitting parameter.
The obtained values are then used in the ETDS at $\vartheta = 82.0^{\circ}$,
shown in Fig. \ref{fig_PTFE_sezioni} (see open symbols and their fitting curves).
It is worth noting that, given its rotational diffusive behavior,
the VH component has a constant characteristic decay time, for both wave vectors $Q$
(see Eq. (\ref{eq_tau_VH}) ).
By contrast, the VV component shows a faster decay at larger wave vectors $Q$,
reflecting its translational diffusive behavior (see Eq. (\ref{eq_tau_VV})).

\begin{figure}
\begin{center}
\includegraphics{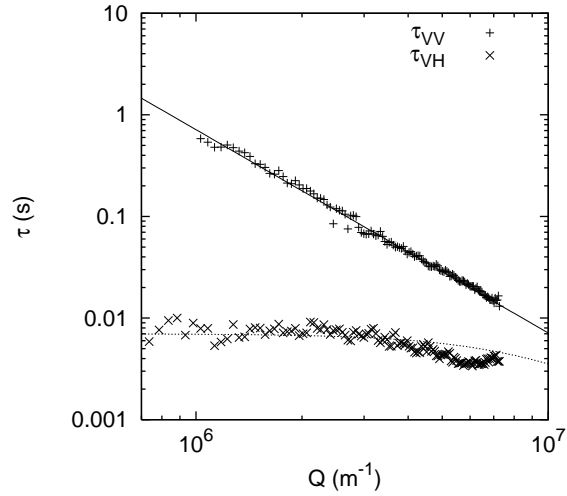}
\end{center}
\caption{Decay times $\tau_{VV}$ and $\tau_{VH}$, obtained by the described fitting procedure,
and plotted as functions of the scattering wave vector $Q$.
Lines represent a fitting on $\tau_{VV}$ and $\tau_{VH}$ 
with Eq. (\ref{eq_tau_VV}) and Eq. (\ref{eq_tau_VH}).
Values obtained by the fit $D_T=1.4\cdot 10^{-12}\mathrm{m}^2/\mathrm{s}$ and $\Theta=23.8\mathrm{s}^{-1}$.
Data taken for PTFE as in Fig. \ref{fig_immagini_pol}.
}
\label{fig_PTFE_decay_times}
\end{figure}

The values of $\tau_{VV}$ and $\tau_{VH}$, obtained by the fitting procedure described above,
are shown for several wave vectors $Q$ in Fig. \ref{fig_PTFE_decay_times}.
At the end, by fitting the resulting data with Eq. (\ref{eq_tau_VV})
and Eq. (\ref{eq_tau_VH}), the translational and rotational diffusion coefficients
of the sample are obtained. The outcomes of this fitting procedure for PTFE sample are
$D_T=1.4\cdot 10^{-12}\mathrm{m}^2/\mathrm{s}$; $\Theta=23.8\mathrm{s}^{-1}$, 
in very good agreement with the values obtained with the traditional dynamic SIFF apparatus.

\section{DR1 experimental results and discussion}

The analysis and the discussion of depolarized SINF measurements for the DR1 sample
follows the one used for PTFE. Namely, we collect the images 
with parallel and nearly crossed polarizer/analyzer, and then we calculate the
power spectra $S\left(q,\Delta t\right)$ at different exposure times $\Delta t$.

\begin{figure}
\begin{center}
\includegraphics{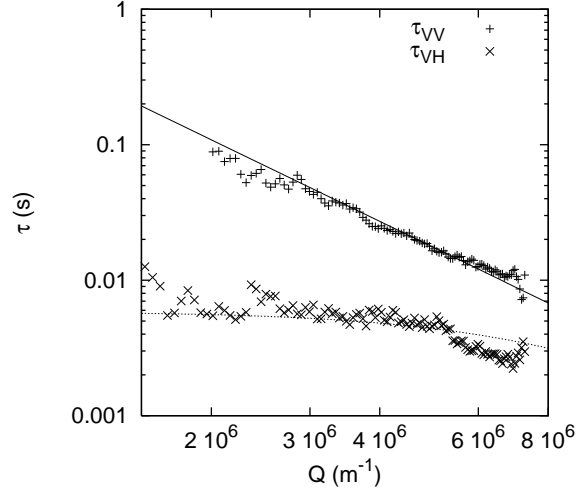}
\end{center}
\caption{Decay times and fitting lines as for Fig. \ref{fig_PTFE_decay_times}. Values obtained by the fit:
$D_T=2.3\cdot 10^{-12}\mathrm{m}^2/\mathrm{s}$; $\Theta=28.2\mathrm{s}^{-1}$, $\alpha=0.477$.
Data taken for the DR1 sample in water, at 1\% volume fraction.
}
\label{fig_DR1_decay_times}
\end{figure}

The $\vartheta=0$ ETDS $S^{VV}\left(q,\Delta t\right)$ 
are simply fitted with Eq. (\ref{eq_ETDS_esponenziale}).
The characteristic times $\tau_{VV}$ obtained with the fitting procedure
are shown in Fig. \ref{fig_DR1_decay_times}.

The VH component is evaluated using Eq. (\ref{eq_S_VH_esplicita}).
In this case the analysis is somewhat different, because the field autocorrelation
function is the stretched exponential described by Eq. (\ref{eq_stretched_exponential}). 
The ETDS can be calculated through Eq. (\ref{eq_ETDS}), and the outcome is:
\begin{equation}
S^{VH}\left(Q,\Delta t\right) \propto f_{\alpha}\left(\frac{\Delta t}{\tau_{VH}}\right)
\label{eq_ETDS_stretched_exponential}
\end{equation}
where:
\begin{equation}
f_{\alpha}\left(x\right) = \frac{2}{\alpha x} 
\left[
\gamma\left(\frac{1}{\alpha}, x^{\alpha}\right) - 
\frac{1}{x}\gamma\left(\frac{2}{\alpha}, x^{\alpha}\right)
\right],
\label{eq_ETDS_forma_stretched}
\end{equation}
and $\gamma\left(a,x\right)$ is the lower incomplete gamma function:
\begin{equation}
\gamma \left(a, x\right) = \int^x_0 e^{-t} t^{a-1} \mathrm{d}t
\end{equation}
For $\alpha=1$, the function $f_{\alpha}\left(x\right)$ reduces to the already described $f\left(x\right)$
shown in Eq. (\ref{eq_ETDS_forma}).

The characteristic times $\tau_{VH}$, obtained by fitting $S^{VH}\left(Q,\Delta t\right)$
with Eq. (\ref{eq_ETDS_stretched_exponential}), are shown in Fig. \ref{fig_DR1_decay_times}.

From the VV term, we get $D_T = 2.3\cdot 10^{-12}\mathrm{m}^/\mathrm{s}$. From the VH term, 
we get $\Theta=28.2\mathrm{s}^{-1}$; $\alpha=0.477$. All the SINF fitting results are very close
to the SIFF results.

\begin{figure}
\includegraphics{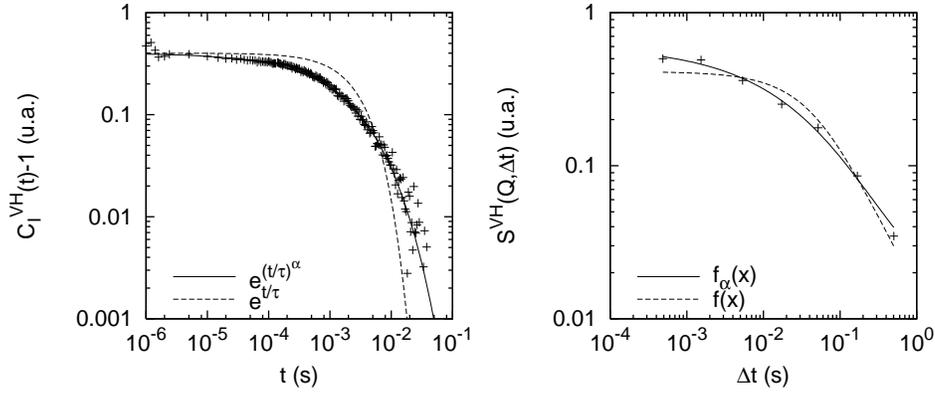}
\caption{
Left panel: depolarized VH intensity autocorrelation function $C_I^{VH}\left(t\right)$ 
measured as a function of 
delay time $t$, by using traditional SIFF apparatus, at $Q=2.5\cdot10^6\mathrm{m}^{-1}$. 
The lines represent a fit with a stretched exponential or a single exponential decay.
Fitting parameters: $\Theta=27.8\mathrm{s}^{-1}$; $\alpha=0.523$.
Right panel: depolarized VH ETDS measured as a function of exposure time $\Delta t$, by using SINF apparatus,
at $Q=2.1\cdot10^6\mathrm{m}^{-1}$.
The lines represent a fit with $f_{\alpha}\left(x\right)$ from 
Eq. (\ref{eq_ETDS_forma_stretched}) (which is related to a stretched exponential),
or a fit with $f\left(x\right)$ from Eq. (\ref{eq_ETDS_forma}) (which is related to a single exponential).
Fitting parameters: $\Theta=28.2\mathrm{s}^{-1}$, $\alpha=0.477$.
Data taken for DR1 sample as in Fig. \ref{fig_DR1_decay_times}.
}
\label{fig_DR1_stretched_exp}
\end{figure}

Figure \ref{fig_DR1_stretched_exp} shows the comparison between the VH intensity
autocorrelation function measured with SIFF, and the VH ETDS measured with SINF,
evaluated with Eq. (\ref{eq_S_VH_esplicita}).
As it appears from the figure, the stretched exponential decay is clearly demonstrated by both
techniques.

\section{Conclusions}

In this paper we present original data obtained with a polarized and depolarized SINF instrument,
here described for the first time. By using the proposed technique, the translational and
rotational diffusion coefficients of generic rod-like colloidal particles can be measured.
The procedure has been validated with PTFE and iron oxide rod-like colloidal samples.
The obtained diffusion coefficients of the rods are in very good agreement with the data acquired with a
traditional dynamic SIFF apparatus. The proposed method allows a simultaneous measurement of the diffusion
coefficients on a very large wave vector range, opening the way to detailed studies of
passive and active motions in anisotropic colloidal systems, ranging from nanoparticles to biological
entities such as viruses, bacteria, proteins, and macromolecules.

\section*{Acknowledgments}
This work has been supported with financial funding from the EU (projects BONSAI 
LSHB-CT-2006-037639 and NAD CP-IP 212043-2).
We thank Solvay Solexis (Bollate, Italy) for the gift of PTFE samples, 
and C. Lancellotti for useful discussions.

\end{document}